# TESTING THE DAE LLRF SYSTEM WITH A PIP-II SSR2 CAVITY

Radhika Nasery[1], Philip Varghese[2], Shrividhyaa Sankar Raman[2], Matei Guran[2], Lennon Reyes[2], Alok Agashe[1],

[1] Bhabha Atomic Research Center, Trombay, Mumbai, India
[2] Fermi National Accelerator Laboratory (FNAL), Batavia, IL 60510, USA

## Abstract

The PIP-II linac is an international collaboration project with in kind contributions of key subsystems from multiple countries including India (DAE). In the research and development phase of the project, the LLRF and resonance control systems were jointly developed by BARC and Fermilab and were delivered to Fermilab for testing and validation. Initial testing of the LLRF system was carried out using Fermilab's analog cavity emulator. Following successful emulator testing, the LLRF system was deployed at STC on a PIP-II 325 MHz SSR2 cavity. The cavity was operated in both SEL and GDR modes at a gradient of 5 MV/m. The results of the testing are presented here.

## INTRODUCTION

Department of Atomic Energy (DAE) India, Institutions, viz, Bhabha Atomic Research Center (BARC), Raja Ramanna Centre for Advanced Technology (RRCAT), Variable Energy Cyclotron Centre(VECC) and the Inter University Accelerator Centre(IUAC) are collaborating with Fermilab National Accelerator Laboratory (FNAL), USA for the design and development of several key technology components in the important area of superconducting RF accelerators under Indian Institutes Fermilab Collaboration (IIFC). Several of these technologies viz, the superconducting magnets, Solid State RF amplifiers, radio-frequency protection and interlock systems, the Low Level RF (LLRF) controls and Resonance Control Systems (RCS) are being designed and engineered at the BARC. In the R&D Phase of the IIFC project, LLRF and RCS systems jointly developed by BARC and FNAL were delivered to FNAL for testing and validation.

## LLRF SYSTEM

The DAE Low-Level Radio Frequency (LLRF) system comprises several key components [1]: a down-converter module, an up-converter module, a digital cavity controller, and a reference phase generator system as shown in Figure 1. The reference phase generator system supplies stable RF signals, including the clock for the digital cavity controller, a phase reference signal, and RF signals used for frequency down- and up-conversions. The down-converter module translates the cavity frequency to a common intermediate frequency (IF), facilitating uniform digital processing across channels. Conversely, the up-converter module performs the reverse operation—it translates the processed intermediate frequency signals back to the cavity frequency, enabling precise control of the accelerating field within the cavity.

At the core of the LLRF system is the digital controller [2], which includes a digitizer board equipped with high-speed Analog-to-Digital Converters (ADCs), Digital-to-Analog Converters (DACs), a precision clock distribution network, and the System-on-Chip Field Programmable Gate Array (SoC FPGA). SoC-FPGA integrated platform combines a high-performance FPGA fabric with an embedded ARM processor, enabling real-time RF signal processing and data acquisition alongside high-level system control and diagnostics. The ADCs digitize the down-converted IF signals for processing, while the DACs generate the control and drive signals required for RF feedback and feed-forward control loops. The FPGA fabric handles time-critical operations such as in-phase/quadrature-phase (I/Q) demodulation, vector sum computation, and fast feedback control. The embedded processor manages tasks including configuration, data logging, and communication with the control infrastructure.

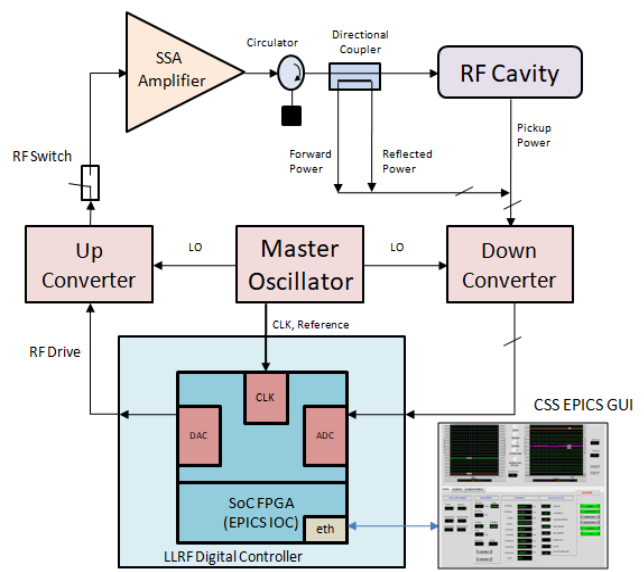

Figure 1: LLRF System

The entire system operates within the Experimental Physics and Industrial Control System (EPICS) framework [3], with the digital controller implemented as an Input/Output Controller (IOC). This architecture supports modular development, real-time feedback, remote diagnostics through EPICS based data acquisition scheme. The operator interface is provided through CSS based EPICS GUI.

## TESTING WITH FNAL CAVITY EMULATOR

Initial validation of the LLRF system was carried out using Fermilab's analog cavity emulator [4], which replicates the dynamic behaviour of superconducting RF cavities. Figure 2 shows the setup of LLRF system with FNAL cavity emulator.

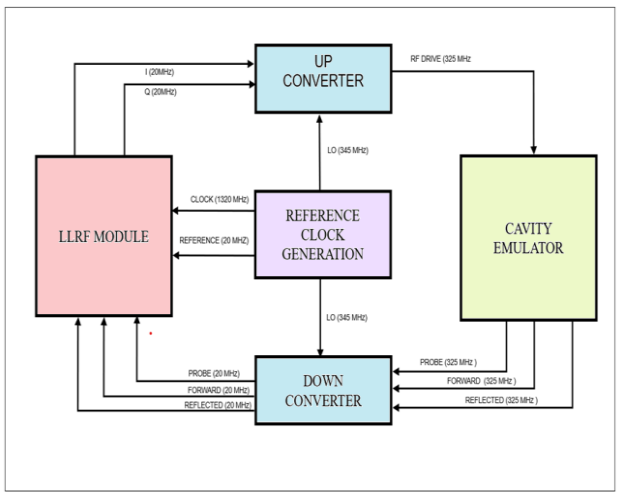

Figure 2: LLRF Setup with FNAL cavity emulator

The emulator provided a controlled and reproducible platform for LLRF algorithm development, system tuning, and stability analysis without the risks and constraints of actual hardware. A thorough verification of the system's operating modes was conducted under simulated conditions, covering both power delivery modes—continuous power (CW) and pulsed power (PL) operation—as well as resonance control modes, including Generator-Driven Resonance (GDR) and Self-Excited Loop (SEL). The evaluation included testing the performance of the feedback loop, gain settings, and phase tracking accuracy. These tests ensured robust and stable cavity field regulation across all expected operational scenarios, demonstrating the system's reliability and adaptability in varying environments. Figure 3 shows the EPICS GUI of LLRF system testing with FNAL Cavity Emulator in Pulsed mode. This significantly accelerated the development cycle and ensured robust performance before integration with real SRF systems.

## TESTING AT SPOKE RESONATOR TEST CAVE (STC):

The STC at Fermilab has been developed for the characterization of 325 MHz Single Spoke Resonator cavities (SSR1 and SSR2) [5] integrated with high-power RF couplers. A Low-Level RF (LLRF) control system is employed at STC for both cavity characterization and field regulation, ensuring optimal RF performance under varying operational conditions. A Resonance Control

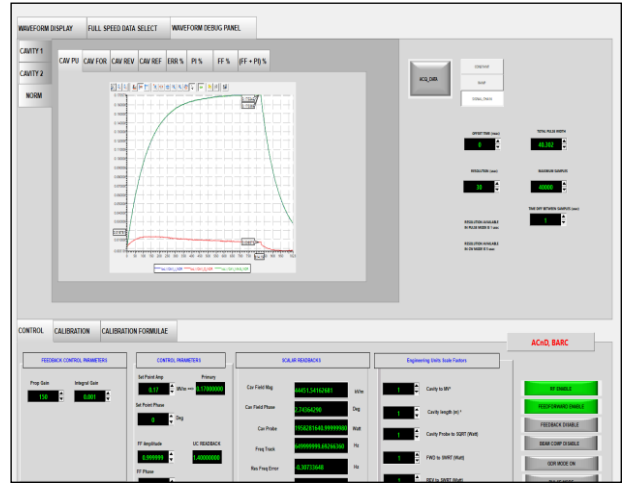

Figure 3: FNAL cavity emulator testing results in pulse mode.

System (RCS) is used in conjunction to manage cavity frequency tuning.

Following successful emulator testing, the DAE LLRF system's LLRF Digital module, up-converter module and down-converter module was deployed at STC on a 325 MHz SSR2 PIP-II superconducting RF (SRF) cavity. The SSR2 cavity under test exhibited a loaded quality factor (Q) of $4.9751 \times 10^6$. Figure 4 shows the LLRF system installed at STC facility of Fermilab.

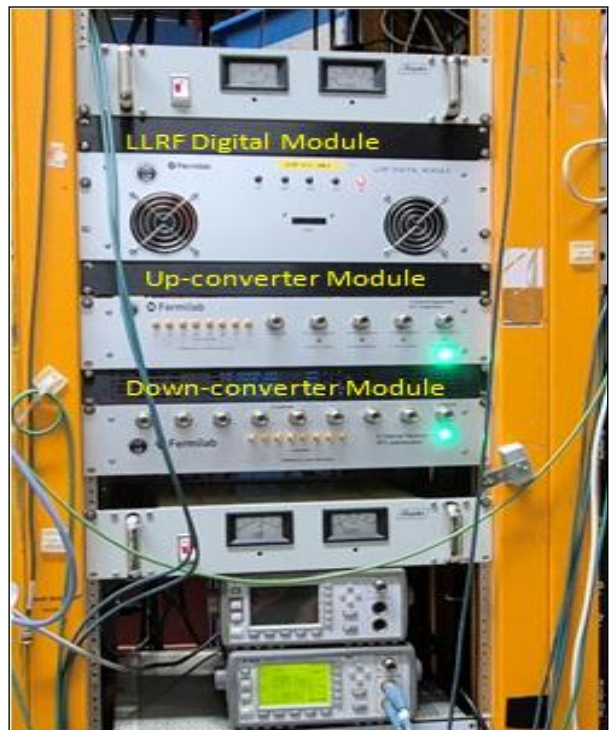

Figure 4: DAE LLRF system installed at STC

*Testing at STC: Calibrations*

The LLRF system EPICS GUI provides programmable calibration parameters. The system was calibrated such that the forward power, reflected power, and cavity accelerating gradient readings displayed on the GUI matched those obtained from a calibrated external power meter. These values were further cross-verified against Fermilab's LLRF system to ensure consistency and accuracy. Figure 5 shows a comparison between the power meter measurements and the corresponding GUI readings when the SSR2 cavity at STC is at 5 MV/m.

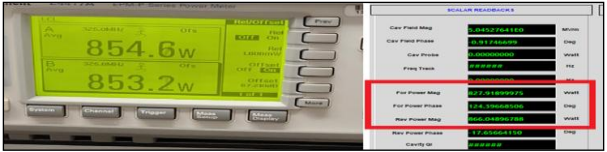

Figure 5: DAE LLRF system installed at STC

*Testing at STC: SEL*

The SEL mode is a critical operational mode for superconducting RF cavities, particularly during initial commissioning and tuning. Testing at STC was started with SEL mode as it allows the RF drive to automatically track and lock to the cavity's natural resonance frequency, compensating for detuning caused by mechanical vibrations or thermal effects. This ensured stable field build-up without requiring prior knowledge of the exact cavity resonance frequency.

In continuous wave (CW) SEL mode, the LLRF system successfully tracked the cavity resonance; while a stepper motor tuner was used to align the resonance frequency near the nominal 325 MHz. The SEL operation was successfully carried out with active feedback, while maintaining a cavity gradient of 5 MV/m with a Forward Power (FP) of 861.6 Watts, as demonstrated in the EPICS GUI shown in Figure 6. Figure 7 shows the EPICS GUI display, demonstrating that the LLRF system is able to track the cavity resonance frequency.

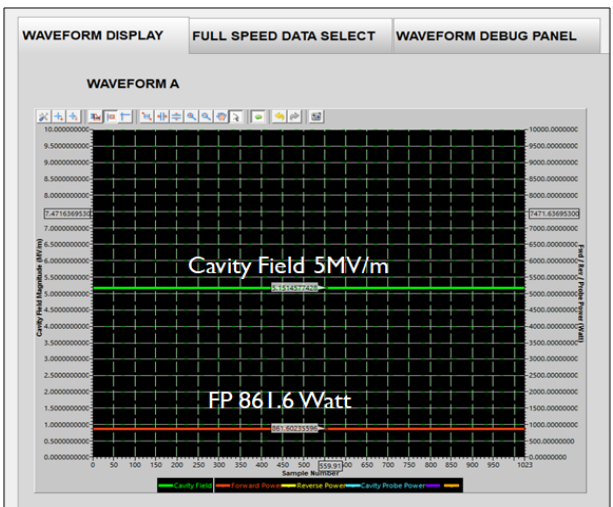

Figure 6: LLRF CW SEL operation: Cavity Field and FP

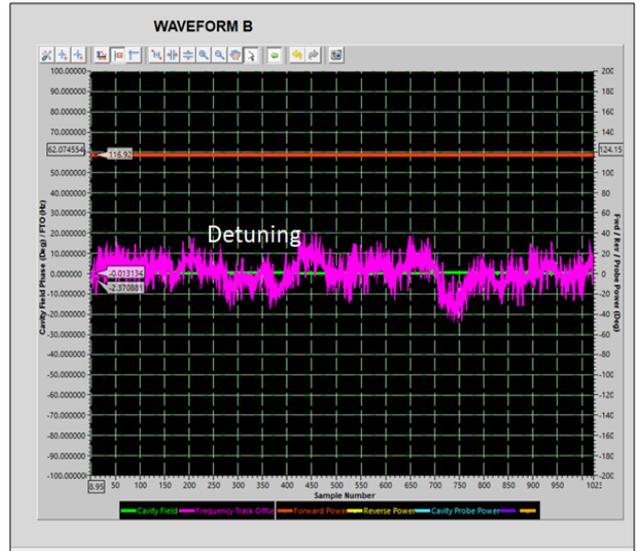

Figure 8: LLRF CW SEL operation: Detuning

SEL mode operation was also tested in pulsed mode (PL) with a pulse rate of 20 Hz and on time with 50% duty cycle. Figure 9 shows the SEL operation in PL mode with cavity field rising till 5MV/m.

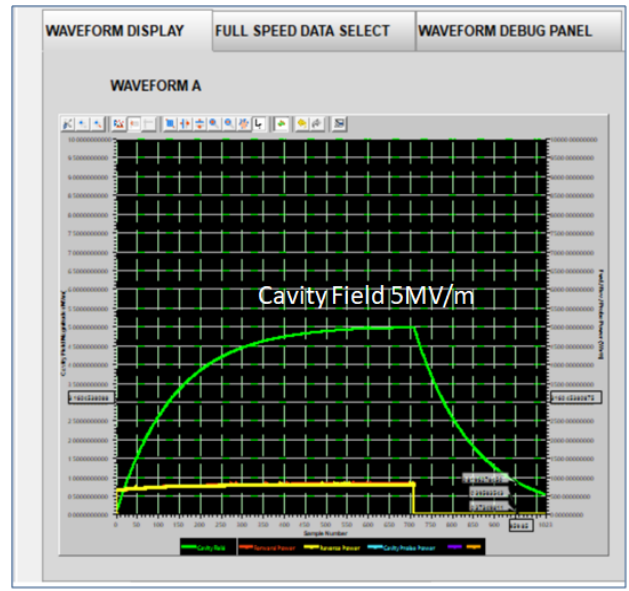

Figure 9: LLRF PL SEL operation: Cavity Field

*Testing at STC: GDR*

In the GDR mode, the cavity is driven at a fixed frequency (325 MHz for SSR2 cavity), phase-locked to an external stable reference. GDR mode is essential as it drives all cavities using a common external stable reference, ensuring synchronized operation and consistent phase coherence between RF cavities.

CW GDR mode was operated with active feedback control and LLRF system was successfully able to maintain a cavity gradient of 5MV/m. Figure 10 shows EPICS GUI of Cavity Field gradient in CW mode.

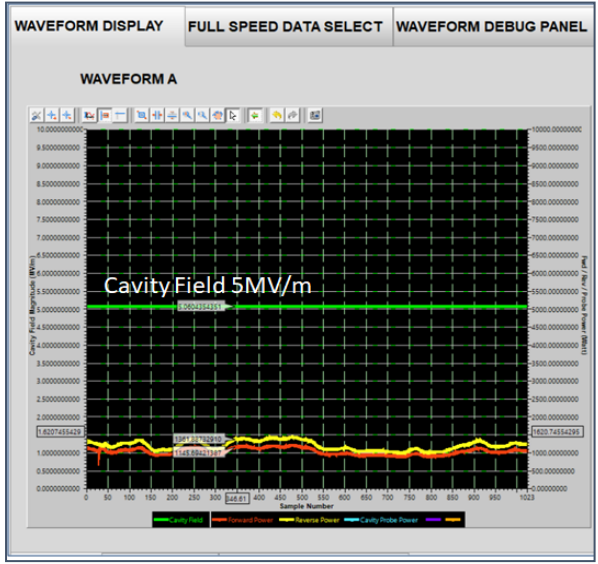

Figure 10: LLRF CW GDR operation: Cavity Field

*Testing Resonance Control System (RCS)*

The EPICS-based RCS developed under IIFC has been tested in a lab setup at FNAL using the Phytron stepper motor, which is used in the tuners for all types of cavities in PIP-II. RCS operation with limit switches was also verified (Figure 11). Piezo was tested in manual mode and its voltage ramp operation was tested.

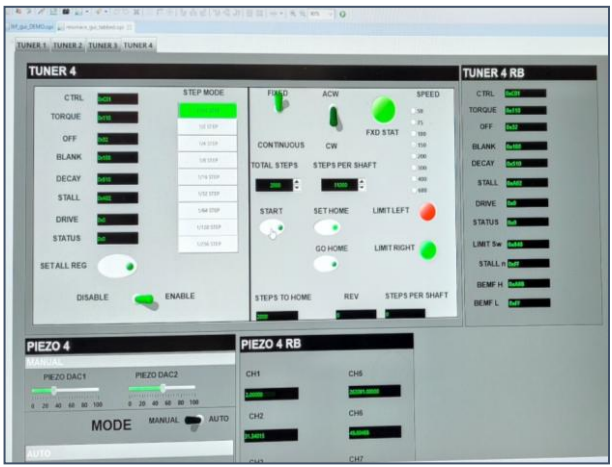

Figure 11: RCS EPICS GUI showing Left LIMIT SW pressed

## CONCLUSION AND FUTURE WORK

The system achieved stable field gradients up to 5 MV/m in CW SEL mode, both with and without active feedback. Additional testing in 20 Hz pulsed mode (50% duty cycle, no feedback) at the same gradient provided insight into system behaviour under pulsed conditions. In GDR mode with feedback engaged, the LLRF system maintained a stable cavity gradient of 5 MV/m, demonstrating robust performance under high-power operation. Consistent results were obtained using Fermilab's LLRF system, confirming its reliability.

## ACKNOWLEDGEMENT

The authors would like to express their sincere gratitude to the BARC team for providing invaluable support and contributions throughout the development and operation phases of this work. In particular, we acknowledge the efforts of Paresh Motiwala, Nibandh Kumar and C.I. Sujo, for their expertise and continued efforts in the development and maintenance of the LLRF system and Sandeep Bharade for embedded EPICS platform development. We also acknowledge the support of the Shri. M.Y. Dixit [Head AEAS] and Shri. U.D. Malshe [Director MRG] whose guidance, encouragement, and leadership were pivotal throughout this project. Their insights and continuous support have greatly contributed to the successful completion of this work. We would also like to thank the LLRF team at Fermilab for their support and guidance in completing these tests.